\documentclass[aps,prl,twocolumn,preprintnumbers,linenumber,amsmath,amssymb,superscriptaddress]{revtex4-1}
\usepackage{graphicx}
\usepackage{subfigure}
\usepackage{epsfig}
\usepackage{dcolumn}
\usepackage{bm}
\usepackage{ulem}
\usepackage{dcolumn}
\usepackage{color}
\usepackage{amsbsy}
\usepackage{lineno}
\usepackage{blindtext}

\def\avg#1{\left\langle#1\right\rangle}
\def\bra#1{\left\langle#1\right|}
\def\ket#1{\left|#1\right\rangle}

\def\be{\begin{equation}}       \def\ee{\end{equation}}
\def\bea{\begin{eqnarray}}      \def\eea{\end{eqnarray}}
\def\ba{\begin{array} }
\def\ea{\end{array} }
\def\bnum{\begin{enumerate} }
\def\enum{\end{enumerate}}

\def\=>{\Rightarrow}
\def\>{\rightarrow}
\def\A{\uparrow}
\def\V{\downarrow}

\def\eye2{Fathbb{I}}

\def\Eq#1{Eq.~(\ref{#1})}
\def\Fig#1{Fig.~\ref{#1}}

\def\Tr{\mathrm{Tr}}

\renewcommand{\>}{\rangle}

\begin{document}

\title{Majorana-Time-Reversal Symmetries: \\A Fundamental Principle for Sign-Problem-Free Quantum Monte Carlo Simulations}
\author{Zi-Xiang Li}
\affiliation{Institute for Advanced Study, Tsinghua University, Beijing 100084, China}
\author{Yi-Fan Jiang}
\affiliation{Institute for Advanced Study, Tsinghua University, Beijing 100084, China}
\affiliation{Department of Physics, Stanford University, Stanford, California 94305, USA}
\author{Hong Yao}
\email{yaohong@tsinghua.edu.cn}
\affiliation{Institute for Advanced Study, Tsinghua University, Beijing 100084, China}

\begin{abstract}
A fundamental open issue in physics is whether and how the fermion sign problem in quantum Monte Carlo (QMC) simulations can be solved generically. Here, we show that Majorana-time-reversal (MTR) symmetries can provide a unifying principle to solve the fermion sign problem in interacting fermionic models. By systematically classifying Majorana-bilinear operators according to the anti-commuting MTR symmetries they respect, we rigorously proved that there are two and only two \textit{fundamental} symmetry classes which are sign-problem-free and which we call the ``Majorana class'' and ``Kramers class'', respectively. Novel sign-problem-free models in the Majorana class include interacting topological superconductors and interacting models of charge-4e superconductors. We believe that our MTR unifying principle could shed new light on sign-problem-free QMC simulation on strongly correlated systems and interacting topological matters.
\end{abstract}
\date{\today}
\maketitle

Interactions between particles are ubiquitous, and studying interacting models of many-body systems is of central importance in modern condensed matter physics\cite{Fradkinbook, XGWenbook, sachdevbook}, quantum chromodynamics (QCD), and other fields. However, almost all interacting models in two and three dimensions, especially those with strong correlations, are beyond the solvability of any known analytical methods. Consequently, developing efficient and unbiased numerical methods plays a key role in understanding many-body physics in solid state materials like high-temperature superconductors \cite{kivelson-rmp,xiaogang-rmp} and other systems such as quark matter. Quantum Monte Carlo (QMC) is among the most important approaches to study interacting many-body systems\cite{Blankenbecler-81,Hirsch-81,Hirsch-85,HQLin-88,Sugiyama-86,Sorella-89,White-89, Ceperley-86, Suzuki-92, Shiwei-95, Prokofev-98, Sandvik-99, Foulkes-01, Assaad-08, Gull-11, Wiese-99}, as it is numerically-exact and intrinsically-unbiased. Nonetheless, QMC often encounters the notorious fermion-sign-problem, 
making it practically infeasible to study those models with large sizes and at a low temperature\cite{Loh-90}. It has been highly desired to find solutions to the fermion-sign-problem in interesting models that are relevant to intriguing systems such as high-temperature superconductors\cite{Zaanen-08}.

\begin{table}[t]
\includegraphics[width=7.9cm]{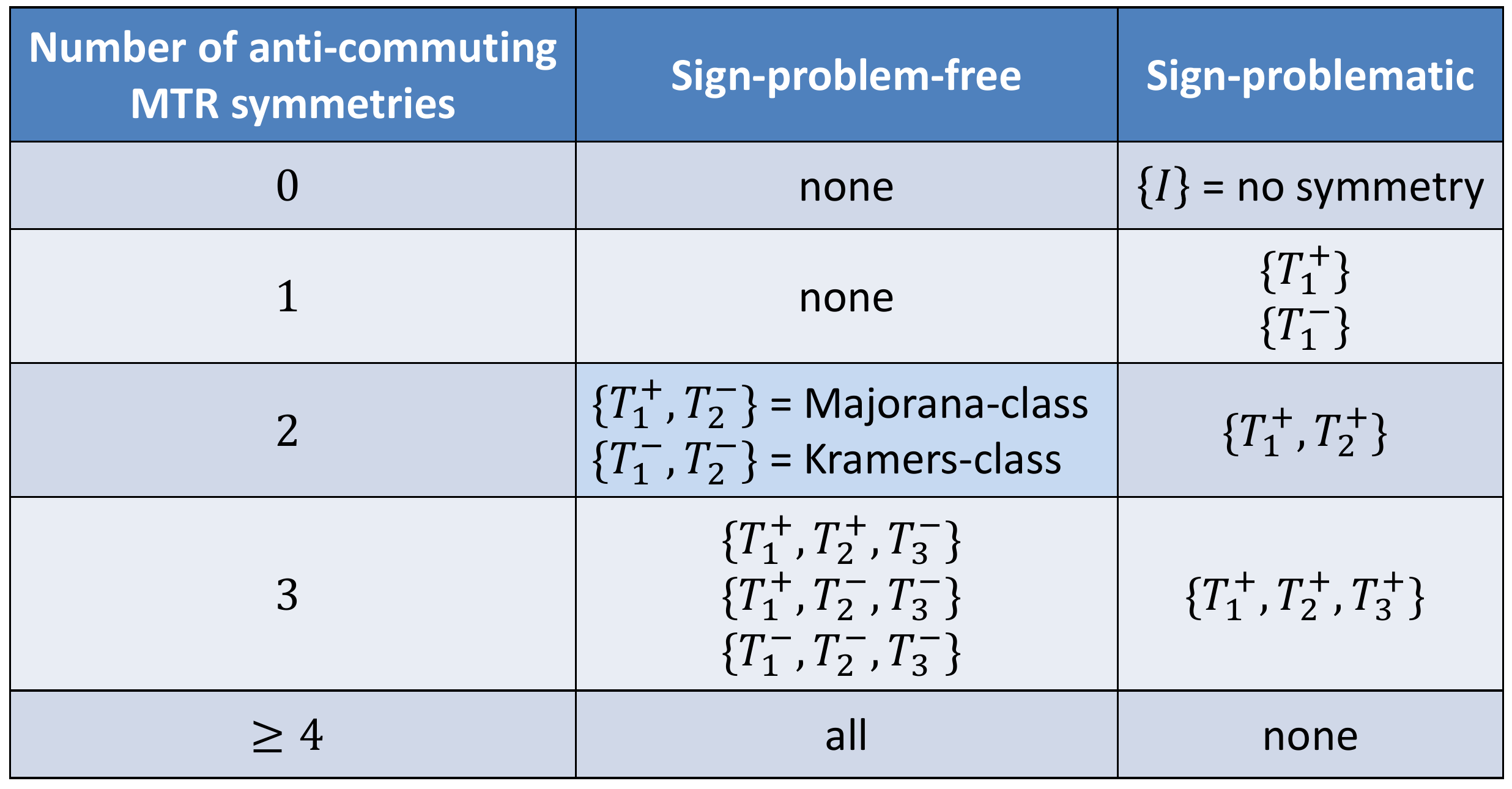}
\caption{The ``periodic table'' of sign-problem-free symmetry classes defined by the set of anticommuting Majorana-time-reversal symmetries $\{T_1^{p_1},T_2^{p_2},\cdots,T_n^{p_n}\}$ they respect, where $p_i=\pm$ and $(T_i^\pm)^2=\pm 1$. We rigorously proved that there are two and only two {\it fundamental} symmetry classes which are sign-problem-free: the Majorana class and Kramers class, respectively. For the former, Majorana-bilinear operators possess two MTR symmetries $T_1^{+}$ and $T_2^{-}$ with $T_1^+T_2^-=-T_2^-T_1^+$, and it is a {\it genuinely new} sign-problem-free symmetry class introduced in Ref. \cite{LJY-15a}, qualitatively different from the latter, which is based on the conventional Kramers-time-reversal symmetry studied in Ref. \cite{Congjun-05}.}
\label{class}
\end{table}

Even though a general solution of the fermion-sign-problem is nondeterministic polynomial (NP) hard\cite{Troyer-05}, many specific interacting models have been successfully identified to be sign-problem-free. 
One prototype sign-problem-free example is the repulsive Hubbard model at half filling\cite{Hirsch-85}.  In the language of auxiliary-field QMC\cite{Blankenbecler-81,Hirsch-81,Hirsch-85}, the partition function $Z=\sum \rho$, where the Boltzmann weight $\rho=\Tr\prod_{i=1}^{N_\tau} \exp[\hat h_i]$ with $\hat h_i$ being fermion-bilinear operators depending on auxiliary fields at imaginary time $\tau_i$. If all Boltzmann weight $\rho>0$, the simulation is free from the fermion sign problem and the needed computation time grows only polynomially with the system size. Tremendous effort has been devoted to construct a fundamental principle for solving the fermion sign problem.

One successful strategy of solving the sign problem is to employ the Kramers symmetry of fermion-bilinear operators $\hat h_i$, which is defined as having both time-reversal symmetry $\hat \Theta$ with $\hat \Theta^2=-1$ and charge conservation $\hat Q$.  With the Kramers symmetry, eigenvalues always appear in Kramers pairs such that the Boltzmann weight can be shown to be positive definite\cite{Congjun-05}. Sign-problem-free models with Kramers-symmetry have been studied extensively during the past three decades. One naturally asks if a more fundamental symmetry principle exists for solving the fermion sign problem in models whose sign-problem solutions remain unknown so far.

Recently, Majorana representation was first introduced by three of us in Ref. \cite{LJY-15a} to solve the fermion sign problem in models (including spinless and spinful fermion models) which are beyond the Kramers method. Here we employ time-reversal symmetry in Majorana representation\cite{LJY-15a} as a fundamental and unifying principle to solve the fermion sign problem. We first classify Majorana-bilinear operators $\hat h=\gamma^T h \gamma$ according to their Majorana-time-reversal (MTR) symmetries, where $h^T=-h$ is an antisymmetric matrix and  $\gamma^T=(\gamma_1,\cdots,\gamma_{2N})$ are Majorana operators with $\{\gamma_i,\gamma_j\}=2\delta_{ij}$\cite{beenakker-13}, and then identify all symmetry classes which {\it must} be sign-problem-free. Note that ``time-reversal'' here generally represents ``antiunitary''. Because Majorana operators are real, Majorana-time-reversal transformation can be represented by $T=UK$, where $U$ are \textit{real} orthogonal matrices and $K$ is complex conjugation with $T^2=\pm 1$ for $U^T=\pm U$. By systematically classifying Majorana-bilinear operators according to the maximal set of anticommuting MTR symmetries they respect, we prove that there are only two \textit{fundamental} symmetry classes of models which are sign-problem-free: the Majorana class and Kramers class, respectively. Other sign-problem-free symmetry classes have higher symmetries than the two fundamental ones, as shown in Table I.

For the Majorana class, the Majorana-bilinear operators possess two anticommuting symmetries $T_1^+$ and $T_2^-$, where $(T_i^\pm)^2 =\pm 1$. Majorana-bilinear operators in this class can always be transformed into two decoupled parts which are time-reversal partners to each other such that it is sign-problem-free\cite{LJY-15a}. For the Kramers class, from anticommuting $T^-_1$ and $T^-_2$, the usual Kramers-time-reversal symmetry can be identified so that they are sign-problem-free. Recently, various correlated models in Kramers class were studied by QMC to investigate high-temperature superconductivity near quantum critical points (QCPs) \cite{Berg-12,LWYL-15a,LWYL-15b,SAK,Berg-15a,Berg-15b,Ashvin-15}.

It is worth pointing out that sign-problem-free models in the \textit{genuinely new} Majorana class include interacting topological superconductors with helical Majorana edge states \cite{Yao-13,Qi-13,Ryu-12} and the minimal model for charge-4e superconductors\cite{JLKY-16}. Note that the sign problems of these models are beyond applicability of other known approaches, especially those requiring particle-number conservation\cite{Shailesh-14,Lei-15}. In contrast, the Majorana approach here is general and can be applied to generic models whether the particle number is conserved or not. As an application of our Majorana approach, we have performed large-scale sign-problem-free Majorana QMC simulations on interacting time-reversal-invariant topological $p+ip$ superconductors of spin-1/2 electrons and found that with increasing interactions the system encounters a quantum phase transition from a topological nontrivial superconducting phase to a topologically trivial one by spontaneously breaking time-reversal symmetry\cite{SM}. To the best of our knowledge, it is the first time that a topological quantum phase transition of spontaneous time-reversal symmetry in superconductors can be studied by numerically-exact and intrinsically-unbiased simulations.

{\bf Majorana-time-reversal symmetry classes:} Time-reversal symmetry plays an important role in classifying random matrices as well as topological insulators/superconducutors\cite{Qi-Zhang-11,Hasan-Kane-10,Kitaev-09,Schnyder-08,Qi-Raghu-09}, and in avoiding the fermion sign problem in QMC\cite{Congjun-05}. The Kramers-time-reversal symmetry\cite{Congjun-05} has been a successful guiding principle for sign-problem-free QMC simulations. Nonetheless, it requires the particle-number conservation and is then not the most general time-reversal symmetry one can utilize to prevent fermion-sign-problem\cite{LJY-15a}.
Thus, constructing a more fundamental and generic symmetry principle to avoid the fermion-sign problem is desired.

In Ref. \cite{LJY-15a} we proposed that time-reversal symmetry in Majorana representation can be used to avoid the sign problem in interacting models. Namely, one can employ Majorana fermions to write fermion-bilinear operators: $\hat h (\tau_i)\equiv \hat h_i = \gamma^T h_i \gamma$, where $\gamma^T =(\gamma^1_1,\cdots,\gamma^1_N,\gamma^2_1,\cdots,\gamma^2_N)$ and $h_i$ is a $2N\times 2N$ matrix. In the case that
\bea\label{decouple}
h_i=\left(\begin{array}{cc}
B_i & 0 \\ 0 & B_i^*
\end{array}\right),
\eea
$\rho=\Tr \prod_{i=1}^{N_\tau} \exp[\hat h_i] $ is positive definite because of the Majorana-time-reversal symmetry $T^+=\tau^x K$, under which $\gamma^1_i \to \gamma^2_i$, $\gamma^2_i\to \gamma^1_i$, and $B_i\to B_i^\ast$ \cite{LJY-15a}. Here $\tau^\alpha$ are Pauli matrices acting in the Majorana space ($1,2$). Because no coupling between $\gamma^1$ and $\gamma^2$ exists in $\hat h_i$, tracing over the Hilbert space of $\gamma^1$ and $\gamma^2$ can be done separately and $\rho_2=\rho^\ast_1$ due to the Majorana-time-reversal symmetry such that $\rho =\rho_1\rho_2>0$.

Note that $h_i$ in \Eq{decouple} also respects another Majorana-time-reversal symmetry $T^-=i\tau^y K$, besides $T^+=\tau^x K$. Moreover, $T^-T^+=-T^+T^-$. One naturally asks the following question: Can any Majorana-bilinear operator respecting anti-commuting $T^+$ and $T^-$ symmetries be transformed into the form in \Eq{decouple} such that it is sign-problem-free? The answer is positive, as shown below. This further motivates us to ask another question: Can anticommuting Majorana-time-reversal symmetries provide a fundamental principle to classify Majorana-bilinear operators such that general sufficient conditions for sign-problem-free models can be constructed? Our answer is also positive, as we prove below.

As Majorana fermion operators are real, Majorana-time-reversal symmetry can be represented by $T^\pm=U^\pm K$, where $(T^\pm)^2=\pm 1$ and $U^\pm$ is a real orthogonal matrix satisfying $(U^\pm)^T=\pm U^\pm$. We propose to systematically classify generic Majorana-bilinear operators $\hat h_i$ according to the maximal set of anticommuting MTR symmetries ${\mathcal C}=\{T_1^{p_1},\cdots,T_n^{p_n}\}$ they respect, namely $[T_j^{p_j},h_i]=0$ and $T_i^{p_i}T_j^{p_j}+T_j^{p_j}T_i^{p_i}=p_i2\delta_{ij} $, where $p_i=\pm$. Because of the sign choices of $p_i=\pm$, there are totally $n+1$ distinct symmetry classes for each $n$. For $n=0$, there is only one symmetry class $\{I\}$, which means that no Majorana-time-reversal symmetry can be found for those Majorana-bilinear operators; while for $n=1$ there are two symmetry classes: $\{T_1^+\}$ and $\{T_1^-\}$. For $n=2$ we have three symmetry classes: $\{T_1^+,T_2^+\}$, $\{T_1^+,T_2^-\}$, and $\{T_1^-,T_2^-\}$. Here, we are concerned with only the symmetries of $h_i$; namely we assume that $h_i$ are random matrices except respecting the specified set of anticommuting MTR symmetries. This classification scheme using anti-commuting symmetries is, in spirit, similar to the one employed by Kitaev using the Clifford algebra to classify random matrices and construct the periodic table of topological insulators and superconductors\cite{Kitaev-09}.

Obviously, if a symmetry class ${\mathcal C}$ is sign-problem-free, any higher symmetry class ${\mathcal C'}$ whose symmetries can generate all the symmetries of ${\mathcal C}$ must be sign-problem-free. For instance, the symmetry class $\{T_1^+,T_2^+,T_3^+,T_4^+\}$ is higher than the symmetry class $\{T_1^+,T_2^-\}$, because $T_2^-$ in the latter can be generated from the former by identifying $T_2^-=T_2^+T_3^+T_4^+$. If the former is sign-problem-free, the latter must be sign-problem-free. Consequently, it would be sufficient to derive all the \textit{fundamental} symmetry classes which are sign-problem-free.

{\bf ``Periodic Table'' of fermion-sign-problem:} It was known that fermion-sign problem can appear in the following three symmetry classes: $\{I\}$, $\{T_1^+\}$, and $\{T_1^-\}$, as sign-problematic examples in these three symmetry classes are known. For instance, $\Tr \exp[x\gamma^1\gamma^2] =2\cos x$, which is negative for $x$$\in$$(\frac{\pi}{2},\pi)$, even though the Majorana-bilinear operator $x \gamma^1 \gamma^2$ respects the $T^-_1$ symmetry ($\gamma^1$$\to$$\gamma^2$, $\gamma^2$$\to$$-\gamma^1$, plus complex conjugation). This illustrates that the symmetry of $T^-_1$ cannot guarantee sign-problem-free. Consequently, symmetry itself for these classes $\{I\}$, $\{T^+_1\}$, $\{T^-_1\}$ is not sufficient to guarantee the absence of the sign problem. We then move to symmetry classes with $n=2$ anticommuting symmetries: $\{T_1^+,T_2^+\}$, $\{T_1^+,T_2^-\}$, and $\{T_1^-,T_2^-\}$. The symmetries in the class $\{T_1^+,T_2^+\}$ cannot guarantee sign-problem-free, because there are known examples with fermion-sign problem in this class, as shown explicitly below. How about the other two classes $\{T_1^+,T_2^-\}$ and $\{T_1^-,T_2^-\}$? It turns out these two are fundamental symmetry classes which are sign-problem-free, as we shall prove below.

If Majorana-bilinear operators $\hat h_i$ respect MTR symmetries in one of the two symmetry classes $\{T_1^+,T_2^-\}$ and $\{T_1^-,T_2^-\}$, 
$\rho=\Tr\prod_{i=1}^{N_\tau} \exp[\hat h_i]>0$. These two are only \textit{fundamental} symmetry classes which are sign-problem-free. We shall prove this below for the two symmetry classes separately. We call the former symmetry class as the ``Majorana class'', while the latter one as the ``Kramers class'' for reasons which will be clear later.

\underline{Majorana class}: In the Majorana class, the random matrix $h_i$ respects two Majorana-time-reversal symmetries $T_1^+=U_1^+K$ and $T_2^-=U_2^-K$, where $U^+_1$ is a real-symmetric orthogonal matrix but $U^-_2$ a real antisymmetric orthogonal matrix. From these two time-reversal symmetries, one can construct a unitary symmetry $P=T_1^+T_2^-=U_1^+U_2^-$. It is straightforward to see that $P$ is a real symmetric matrix satisfying $P^2 = 1$. Consequently, the eigenvalues of $P$ are $\pm 1$. As $[P,h_i]=0$, we can use $P$ to block-diagonalize $h_i$.

We denote the eigenvectors of $P$ with eigenvalue $+1$ as $\chi_a$, namely $P \chi_a = \chi_a$, where $a=1,\cdots,N$. Because $P$ is a real-symmetric matrix, $\chi_a$ can be chosen to be real, {\it i.e.} $\chi^{\ast}_a=\chi_a$. Since $T_1^+$ satisfies $\{T_1^+,P\}=0$, $T^+_1\chi_a$ are eigenvectors of $P$ with eigenvalue $-1$. Now, we are ready to use the basis $\tilde\chi = (\chi_1,\cdots,\chi_N, T_1^+ \chi_1,\cdots,T_1^+ \chi_N)$ to block-diagonalize the $2N\times 2N$ matrix $h_i$, as follows:
\bea\label{block}
\tilde\chi^T h_i \tilde\chi =\left(\begin{array}{cc}
B_i & 0 \\ 0 & B_i^*
\end{array}\right).
\eea
Consequently, the Boltzmann weights $\Tr\prod_{i=1}^{N_\tau} \exp[\hat{h}_i]$ are positive definite, as required by the time-reversal symmetry between the two decoupled blocks $B_i$ and $B_i^\ast$. Because charge conservation is not required for this symmetry class of $\{T_1^+,T_2^- \}$, it is a new {\it sign-problem-free} symmetry class, which was first studied in Ref. \cite{LJY-15a}. We call it the Majorana class.

\underline{Kramers class}: From $T_1^-$ and $T_2^-$ symmetries, a unitary symmetry $Q=T_1^-T_2^-$ can be derived. Because $Q$ is antisymmetric, namely $Q^T=-Q$, one can construct a charge operator $\hat Q=\gamma^T (iQ) \gamma$ such that $[\hat Q,\hat h_i]=0$. It is clear that the combination of $T_1^-$ and $Q$ charge conservation in this symmetry class is equivalent to the Kramers-symmetry since $[T_1^-, iQ]=0$ and $(T_1^-)^2=-1$. The absence of the fermion-sign problem has been shown in Ref. \cite{Congjun-05} according to the observation of the Kramers pairs in eigenvalues. Because of the Kramers symmetry, we denote this symmetry class as the ``Kramers class''.

For symmetry classes with $n\ge 3$, it turns out that all of them, except only one class $\{T_1^+,T_2^+,T_3^+\}$, can generate the symmetries in either the Majorana-class or the Kramers-class. For instance, the symmetry class $\{T_1^+,T_2^+,T_3^-\}$ has higher symmetry than $\{T_1^+,T_2^-\}$ and $\{T_1^+,T_2^-,T_3^-\}$ higher than both $\{T_1^+,T_2^-\}$ and $\{T_1^-,T_2^-\}$. The only remaining unclear class is $\{T_1^+,T_2^+,T_3^+\}$, which cannot generate $\{T_1^+,T_2^-\}$ or $\{T_1^-,T_2^-\}$.  Even though the symmetry class $\{T_1^+, T_2^+, T_3^+\}$ has relatively high symmetries, it can still suffer from the fermion-sign-problem as we show below that there exist explicit examples in this symmetry-class which are sign-problematic.

Now we explicitly demonstrate that the two symmetry classes $\{T_1^+,T_2^+\}$ and $\{T_1^+,T_2^+,T_3^+\}$ can be sign-problematic by considering the spin-$\frac{1}{2}$ repulsive Hubbard model away from half-filling as an example. For the repulsive-$U$ Hubbard model
\bea
H = -t \sum_{\langle ij \rangle \sigma } \big[ c^\dagger_{i\sigma} c_{j\sigma}  + h.c. \big] \!+\! U\sum_{i} n_{i\uparrow} n_{i\downarrow} \!-\!\mu \sum_{i\sigma} n_{i\sigma},~~
\eea
where $U$$>$$0$, $\mu$$\neq$$0$, and $\sigma$=$\A,\V$, we obtain the following decoupled Majorana-bilinear Hamiltonian $\hat h_n$ after utilizing the Majorana representations $c_\sigma=(\gamma^1_\sigma+i\gamma^2_\sigma)/2$ and performing a Hubbard-Stratonovich (HS) transformation: $\hat h_n \!=\! -\frac{\tilde t}{2} \sum_{\langle ij \rangle }  \gamma^T_i \sigma^0 \tau^y \gamma_j \!+\!   \sum_{i} \!\big[\frac{\tilde \mu}{4}  \gamma^T_i \sigma^0 \tau^y  \gamma_i  \!-\! \lambda \phi^n_{i}  \gamma^T_i i \sigma^y \tau^z \gamma_i\big]$,
where $\phi^n_i$ are auxiliary fields on site $i$ at imaginary time $\tau_n$, $\gamma^T_i = (\gamma^1_{i\uparrow}, \gamma^1_{i\downarrow}, \gamma^2_{i\uparrow}, \gamma^2_{i\downarrow})$, $\sigma^a$ is Pauli matrix in spin space and $\tau^a$ in Majorana space \cite{SM}. It is straightforward to show that $\hat h_n$ possesses three MTR symmetries $T_1^+ = \sigma^x \tau^x K$, $T_2^+ = \sigma^z \tau^x  K$, and $T_3^+ =  \sigma^0 \tau^z K$. Even though respecting these symmetries, the appearance of sign-problem in this decoupled channel for the doped repulsive Hubbard model is well-known. In order to further confirm it, we have also computed the Boltzmann weights for different auxiliary-field configurations and find that negative weights can indeed appear. Namely, models in the symmetry-class $\{T_1^+,T_2^+,T_3^+\}$ can be sign-problematic. Since the class $\{T_1^+,T_2^+,T_3^+\}$ has higher symmetries than  $\{T_1^+,T_2^+\}$, the latter is also sign-problematic.

Combining the results above, we have shown that there are two and only two \textit{fundamental} sign-problem-free symmetry classes: the Majorana-class and the Kramers-class.  The ``periodic table'' of symmetry classes which are sign-problem-free or sign-problematic is shown in Table \ref{class}. It provides a fundamental principle to identify sign-problem-free interacting fermion models. 

{\bf Interacting topological superconductors in the Majorana-class:} So far novel interaction effects in topological superconductors have not been investigated by large-scale QMC mainly due to the lack of such sign-problem-free models, although interacting topological insulators have been much studied by sign-problem-free QMC\cite{Assaad-11,CJWu-11,Assaad-12,Assaad-13,Fiete-13,Wessel-13,CKXu-15,CKXu-16}.  As mentioned above, there is no requirement of charge-conversation for models in the Majorana-class. Consequently, interesting sign-problem-free models describing interacting superconductors may be identified in this class such that we can study strong correlation effect in superconductors using sign-problem-free QMC. Indeed, we have found interacting models of topological superconductors with helical Majorana edge states which are sign-problem-free in the Majorana-class.

We first consider the Hamiltonian describing a topological superconductor of spin-1/2 electrons on the square lattice with time-reversal symmetry:
\bea\label{tsc}
H \!=\! \sum_{ij,\sigma} \!\big[\!-t_{ij} c^\dagger_{i\sigma} c_{j\sigma} \!+\!\Delta_{ij,\sigma}c^\dagger_{i\sigma}c_{j\sigma}^\dagger \!+\! h.c.\!\big]\! \!-\!U\sum_{i} n_{i\uparrow} n_{i\downarrow},~~~
\eea
where $c^\dag_{i\sigma}$ creates spin-1/2 electrons with spin polarization $\sigma=\uparrow,\downarrow$,  $n_{i\sigma}=c^\dag_{i\sigma}c_{i\sigma}$, $t_{ij}=t$ is nearest-neighbor hopping, and $t_{ii}=\mu$ is the chemical potential. When $\Delta_{ij,\sigma}= \Delta$ for $j=i+\hat x$ and $\Delta_{ij,\sigma}= i\sigma\Delta$ for $j=i+\hat y$, the Hamiltonian in Eq. \eqref{tsc} describes a helical topological superconductor\cite{Schnyder-08,Kitaev-09,Qi-Raghu-09} with ($p$+$ip$) triplet-pairing of spin-up electrons and ($p-$$ip$) triplet-pairing of spin-down electrons, which hosts helical Majorana edge states protected by the Majorana-time-reversal symmetry $T^-=i\sigma^y\tau^zK$, where $\sigma^i$ acts in spin-space and $\tau^i$ in the Majorana-space. Besides $T^-$, it also possesses a unitary symmetry $P=\sigma^z$, such that we can construct another Majorana-time-reversal symmetry $T^+ = P T^- = \sigma^x\tau^z K$. In the complex fermions basis, these two anti-commuting TR symmetries are: $T^+ = \sigma^x K$ and $T^- = i\sigma^y K $ where $\sigma^a$ is Pauli matrix acting in spin space $(c_\uparrow, c_\downarrow)$. Note that the unitary symmetry $P=\sigma^z$ is not a U(1) symmetry conserving the total $S^z$ because the triplet pairing in the Hamiltonian breaks the U(1) symmetry. Instead, it only conserves the spin parity $(-1)^{N_\V}$, where $N_\V$ is the number of spin-down electrons.  The topological classification of superconductors respecting the time-reversal symmetry and the spin-parity symmetry in the non-interacting limit is $\mathbb{Z}$.  In the presence of interactions, its topological classification was shown to be $\mathbb{Z}_8$\cite{Qi-13,Yao-13,Ryu-12}.

\begin{figure}[t]
\includegraphics[width=6.cm]{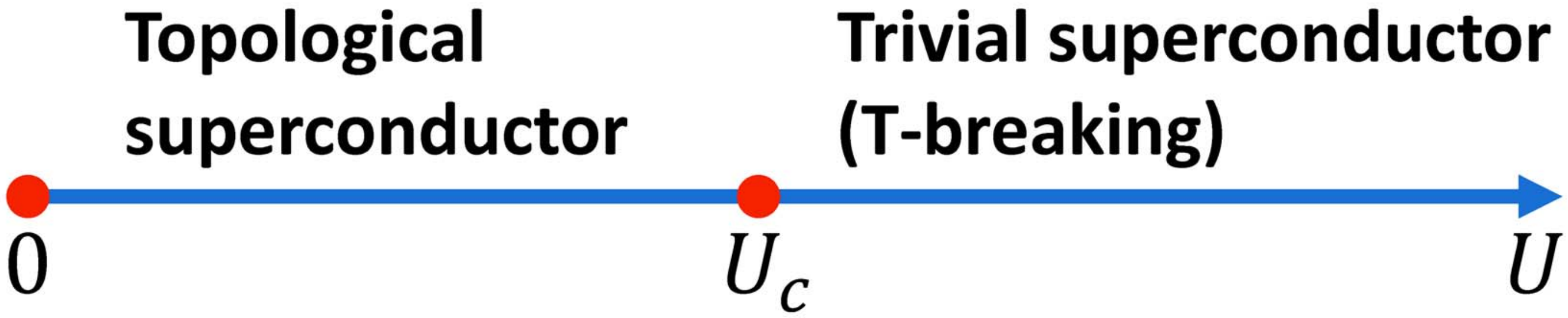}
\caption{The quantum phase diagram of interacting topological superconductors with a topological quantum phase transition as well as the nature of the quantum critical point have been studied by our sign-problem-free QMC simulations \cite{SM}.}
\label{spinful}
\end{figure}

For the Hubbard interactions, we perform the following HS transformation:
\bea
e^{\frac{U}{4} \Delta\tau i \gamma^1_{i\uparrow} \gamma^2_{i\uparrow} i \gamma^1_{i\downarrow} \gamma^2_{i\downarrow}} = \sum_{\phi_i=\pm 1} A e^{ \lambda \phi_i ( i \gamma^1_{i\uparrow} \gamma^2_{i\uparrow} + i \gamma^1_{i\downarrow} \gamma^2_{i\downarrow} )},
\eea
where $\phi_{i}$ represent auxiliary fields living on site $i$, $\Delta\tau$ is the imaginary time slice in the Trotter decomposition, $\lambda=\frac{1}{2}\cosh^{-1}(e^{U\Delta\tau/2})$, and $A=\frac{1}{2}e^{-U\Delta\tau/4}$. It is clear that the decoupled Majorana-bilinear operators also respect both $T^+=\sigma^x\tau^z K$ and $T^-= \sigma^x\tau^z K$ such that this model is sign-problem-free in the Majorana-class. We performed projector QMC \cite{Sorella-89,White-89,Sugiyama-86} simulations which show that, with increasing $U$, the system undergoes a topological quantum phase transition from a topological SC to a trivial SC which breaks time-reversal symmetry spontaneously \cite{SM}, as schematically shown in \Fig{spinful}. The universality class of this topological phase transition is also obtained by our QMC simulations \cite{SM}.  

{\bf Concluding remarks:} We have shown that anti-commuting MTR symmetries can provide a fundamental principle to identify sign-problem-free interacting models. Note that this does not contradict the no-go theorem\cite{Troyer-05} as the symmetry principle introduced here may not directly provide recipes of the sign-problem for all models. Assuming no other requirement than respecting a set of anti-commuting MTR symmetries, we have proved that there are two and only two \textit{fundamental} sign-problem-free symmetry classes.

Here, we focus on the generic symmetry principle for sign-problem-free models. In case that the matrices $h_i$ are not \textit{fully} random besides respecting the required symmetries, it is possible that sign-problem-free models can be found in the symmetry classes $\{I\}$, $\{T_1^+\}$, $\{T_1^-\}$, $\{T_1^+, T_2^+\}$, and $\{T_1^+, T_2^+, T_3^+\}$ \cite{Wiese-99,Congjun-16}. For instance, it was shown in Refs. \cite{Shailesh-14,Congjun-16} that certain models in the symmetry class $\{T_1^+\}$ with special conditions besides the symmetry requirement could be sign-problem-free. In other words, it provides hope to avoid the fermion-sign-problem in various interesting models, such as the repulsive Hubbard model away from half-filling \cite{White-89}, by utilizing a combination of special features and symmetry principles. We believe that the Majorana-time-reversal symmetry principle has shed new light on avoiding fermion-sign-problem in strongly-correlated models describing systems such as high-temperature superconductors and featuring exotic quantum critical phenomena\cite{Isakov-06,Melko-12,FIQCP-15,WAGuo-16}.

{\it Acknowledgement}: We would like to thank Shao-Kai Jian and A. Kitaev for helpful discussions. This work was supported in part by the NSFC under Grant No. 11474175 at Tsinghua University (ZXL,YFJ, and HY).

\begin{widetext}

\renewcommand{\theequation}{S\arabic{equation}}
\setcounter{equation}{0}
\renewcommand{\thefigure}{S\arabic{figure}}
\setcounter{figure}{0}
\renewcommand{\thetable}{S\arabic{table}}
\setcounter{table}{0}

\subsection{I. Sign-problem-free condition for Projector QMC in Majorana class}

Projector QMC 
is an algorithm to explore the ground properties of the systems. In projector QMC, the expectation value of an observable in the ground state can be evaluated as:
\bea
\frac{ \bra{\psi_{0}} O \ket{\psi_0}}{ \avg{\psi_{0} \mid \psi_{0} } }   = \lim_{\theta\rightarrow \infty} \frac{ \bra{\psi_T}e^{-\theta H } O e^{-\theta H} \ket{\psi_T}}{ \bra{\psi_T} e^{-2\theta H} \ket{\psi_T}},
\eea
where $ \psi_0 $ is the true ground state wave function and $ \psi_T$ is a trial wave function which should have a finite overlap with the true ground state wave function. We should emphasize that the finite-temperature proof of sign-problem-free in the Majorana class can be generalized to projector QMC straightforwardly. Similar to the Boltzmann weight in finite-temperature algorithm, the weight for each auxiliary field configuration in the projector algorithm is given by the following expectation value
\bea\label{determinant}
W=\bra{\psi_T} [\prod_{i=1}^{N_\tau} e^{\frac{1}{4}\gamma^T h_i  \gamma} ] \ket{\psi_T},  
\eea
where $\gamma^T=(\gamma^1_1,\cdots,\gamma^1_N,\gamma^2_1,\cdots,\gamma^2_N)$ represents a $2N$-dimensional vector of Majorana fermions and $\ket{\psi_T} = \prod_{a=1}^{2N_f}( \gamma \tilde{P})_a \ket{0}$ is the trial wave function. Here $\tilde P$ is a $2N\times 2N_f$ projector matrix. This projector matrix can be rewritten as $\tilde{P} = \tilde{\chi} \xi $, where $\xi=F\oplus F^\ast$. $F$ is a $N\times N_f$ matrix, which can be written $F=(\phi_1,\cdots,\phi_{N_f})$ with $\phi_i$ an $N$-dimensional vector. $\phi_i$ is often chosen to be single-particle eigenstate of the non-interacting Hamiltonian. In the new Majorana fermion basis $(\alpha,T_1^+\alpha)=\gamma \tilde \chi$, the weight $W$ can be written as
\bea
W=w w^\ast,
\eea
with
\bea
w=\bra{0} (\alpha F)^\dag [\prod_{i=1}^{N_\tau} e^{\frac{1}{4}\alpha^T B_i\alpha} ] (\alpha F)\ket{0}.
\eea
Since $W=ww^\ast$, it is clear that it is positive definite. As $w=\pm [\det(F^\dag\prod_{i=1}^{N_\tau} e^{B_i} F)]^{1/2}$, we obtain
\bea
W=\left|\det(F^\dag\prod_{i=1}^{N_\tau} e^{B_i} F)\right|.
\eea.

\subsection{II. Sign-problematic examples in the symmetry classes $\{T^+_1, T^+_2\}$ and  $\{T^+_1,T^+_2,T^+_3\}$  }


 Even though the symmetry class $\{T_1^+, T_2^+, T_3^+\}$ has relatively high symmetries, it can still suffer from the fermion-sign-problem. We consider spin-$\frac{1}{2}$ repulsive Hubbard model away from half filling as an example. The Hamiltonian is:
\bea
H = -t \sum_{\langle ij \rangle \sigma } \big[ c^\dagger_{i\sigma} c_{j\sigma}  + h.c. \big] \!+\! U\sum_{i} n_{i\uparrow} n_{i\downarrow} \!-\!\mu \sum_{i\sigma} n_{i\sigma},~~
\eea
where $U$$>$$0$, $\mu$$\neq$$0$, and $\sigma$=$\A,\V$. We take basis transformation $c_\sigma=(\gamma^1_\sigma+i\gamma^2_\sigma)/2$ and rewrite the Hamiltonian in Majorana representation:
\bea
 H \!=\! -\frac{t}{2} \sum_{\langle ij \rangle }  \tilde{\gamma}^T_i \sigma^0 \tau^y \tilde{\gamma}_j \!+\!   \sum_{i} \!\big[\frac{ \mu}{4}  \tilde{\gamma}^T_i \sigma^0 \tau^y  \tilde{\gamma}_i  \! + \!  \frac{U}{4} i \gamma_{i\uparrow}^1\gamma^2_{i\uparrow}i\gamma^1_{i\downarrow}\gamma^2_{i\downarrow}     \big]
\eea
where $\tilde{\gamma}^T_i = (\gamma^1_{i\uparrow}, \gamma^1_{i\downarrow}, \gamma^2_{i\uparrow}, \gamma^2_{i\downarrow})$ and $\sigma^a$ is Pauli matrix in spin space and $\tau^a$ in Majorana space. Then we take Hubbard-Stratonovich(HS) transformation to decouple the Hubbard interaction term:
\bea
e^{-\frac{U}{4}\Delta_\tau  i\gamma^1_{i\uparrow}\gamma^2_{i\uparrow} i\gamma^1_{i\downarrow}\gamma^2_{i\downarrow} } =
\frac{1}{2} \sum_{\phi_i = \pm 1} e^{\frac{1}{2}\lambda  \phi_i (\gamma^1_{i\uparrow}\gamma^1_{i\downarrow} - \gamma^2_{i\uparrow}\gamma^2_{i\downarrow}) - \frac{U}{4}\Delta_\tau }
\eea
where $\lambda = \cosh^{-1}(e^{U\Delta_\tau/2})$, $\phi_i$ is auxiliary field living on site $i$ and $\Delta_\tau$ is imaginary time slice under Trotter decomposition. After HS transformation the partition function can be expressed:
\bea
\Tr[e^{-\beta H }] = \sum_{\phi_i^n = \pm 1} \Tr[\prod_n e^{\hat{h}_n(\phi^n)}]
\eea
$\hat{h}_n$ is bilinear Majorana fermions operator at $n$-th imaginary time slice:
\bea
\hat h_n \!=\! -\frac{\tilde t}{2} \sum_{\langle ij \rangle }  \gamma^T_i \sigma^0 \tau^y \gamma_j \!+\!   \sum_{i} \!\big[\frac{\tilde \mu}{4}  \gamma^T_i \sigma^0 \tau^y  \gamma_i  \!-\! \lambda \phi^n_{i}  \gamma^T_i i \sigma^y \tau^z \gamma_i\big]
\eea
where $\tilde{t} = t\Delta_\tau$ and $\tilde{\mu} = \mu \Delta_\tau$. It is straightforward to show that $\hat h_n$ possesses three MTR symmetries $T_1^+ = \sigma^x \tau^x K$, $T_2^+ = \sigma^z \tau^x  K$, and $T_3^+ =  \sigma^0 \tau^z K$. The appearance of sign problem in this decoupled channel of doped Hubbard model is well-known. In order to further confirm it, we have also computed the Boltzmann weights for different auxiliary-field configurations and find that negative weights can indeed appear. This sign-problematic example illustrates that the symmetry class $\{T^+_1,T^+_2,T^+_3\}$ cannot guarantee sign-problem-free. Since $\{T^+_1,T^+_2,T^+_3\}$ has higher symmetries than $\{T^+_1,T^+_2\}$ and $\{T^+_1\}$, both $\{T^+_1,T^+_2\}$ and $\{T^+_1\}$ are also sign-problematic.

\subsection{III. Detailed proof of sign-problem-free for interacting topological superconductors of spin-1/2 electrons}

The topological superconductor of spin-1/2 electrons on the square lattice with attractive Hubbard interactions can be described by the following Hamiltonian:
\bea\label{SM-tsc}
H = \sum_{\langle ij\rangle,\sigma} \big[-t c^\dagger_{i\sigma} c_{j\sigma} +\Delta_{ij,\sigma}c^\dagger_{i\sigma}c_{j\sigma}^\dagger + h.c.\big]-\mu \sum_i (n_{i\uparrow} + n_{i\downarrow})  -U\sum_{i} n_{i\uparrow} n_{i\downarrow},~~~
\eea
where the triplet pairing amplitudes are given by $\Delta_{ij,\sigma}= \Delta$ for $j=i+\hat x$ and $\Delta_{ij,\sigma}= i\sigma\Delta$ for $j=i+\hat y$. We can express complex fermions by two components of Majorana fermions $c_{j\sigma} = \frac{1}{2} ( \gamma^1_{j\sigma} + i \gamma^2_{j\sigma})$, and then rewrite the Hamiltonian as:
\bea
&&H = H_0 + H_I, \nonumber\\
&&H_0 = -\frac{t}{2} \sum_{\avg{ij}}\tilde\gamma^T_i \sigma^0 \tau^y \tilde\gamma_j
+ \frac{i\Delta}{2} \sum_{\avg{ij}_x}\tilde\gamma^T_i \sigma^0 \tau^x \tilde\gamma_j
+ \frac{i\Delta}{2} \sum_{\avg{ij}_y}\tilde\gamma^T_i \sigma^z \tau^z \tilde\gamma_j
+ \frac{\mu}{4} \sum_i \tilde\gamma^T_i \sigma^0 \tau^y \tilde{\gamma_i}, \\
&&H_I =  -\frac{U}{4} \sum_{i} i \gamma^1_{i\uparrow} \gamma^2_{i\uparrow} i \gamma^1_{i\downarrow} \gamma^2_{i\downarrow},
\eea
where $\sigma^\alpha$ and $\tau^\alpha$ are Pauli matrices acting in the spin and Majorana space,  respectively. We can perform Hubbard-Stratonovich transformation of attractive Hubbard term in density channel:
\bea
e^{\frac{U}{4} \Delta\tau i \gamma^1_{i\uparrow} \gamma^2_{i\uparrow} i \gamma^1_{i\downarrow} \gamma^2_{i\downarrow}} = \sum_{\phi_i=\pm 1} A e^{ \lambda \phi_i ( i \gamma^1_{i\uparrow} \gamma^2_{i\uparrow} + i \gamma^1_{i\downarrow} \gamma^2_{i\downarrow} )},
\eea
where $\phi_{i}$ represent auxiliary fields living on site $i$, $\Delta\tau$ is the imaginary time slice in the Trotter decomposition, $\lambda=\frac{1}{2}\cosh^{-1}(e^{U\Delta\tau/2})$, and $A=\frac{1}{2}e^{-U\Delta\tau/4}$. Consequently, the decoupled Hamiltonian after HS transformation is:
\bea
\hat h(\phi) =   \sum_{\avg{ij},\sigma} -\frac{\tilde{t}}{2} \gamma^T_i \sigma^0 \tau^y \gamma_j
+  \sum_{\avg{ij}_x}\frac{i \tilde{\Delta} }{2} \gamma^T_i \sigma^0 \tau^x \gamma_j
+ \sum_{\avg{ij}_y}\frac{i \tilde{\Delta} }{2}  \gamma^T_i \sigma^z \tau^z \gamma_j
 + \frac{\tilde{\mu}}{4} \sum_i \gamma^T_i \sigma^0 \tau^y \gamma_i +  \lambda \sum_i \phi_i \gamma^T_i \sigma^0 \tau^y \gamma_i,
\eea
where $\gamma^T_i=(\gamma^1_{i\uparrow}, \gamma^2_{i\uparrow},\gamma^1_{i\downarrow}, \gamma^2_{i\downarrow})$ and $\tilde{t} = \Delta\tau t, \tilde{\Delta} = \Delta\tau \Delta, \tilde{\mu} = \Delta\tau \mu$ . Because it respects these two anti-commuting MTR symmetries: $T_1^+ = \sigma^x \tau^z K $ and $T_2^- = i\sigma^y \tau^z K $, it belongs to the Majorana-class and is then sign-problem-free according to the theorem we have proved in the main text. In the complex fermions basis, these two anti-commuting TR symmetries are: $T_1^+ = \sigma^x K$ and $T_2^- = i\sigma^y K $ where $\sigma^a$ is Pauli matrix acting in spin space $\tilde c = (c_\uparrow, c_\downarrow)$.

\subsection{IV. Numerical results of QMC simulations of the interacting topological superconductors}

\begin{figure}[t]
\subfigure{\includegraphics[height=4.2cm]{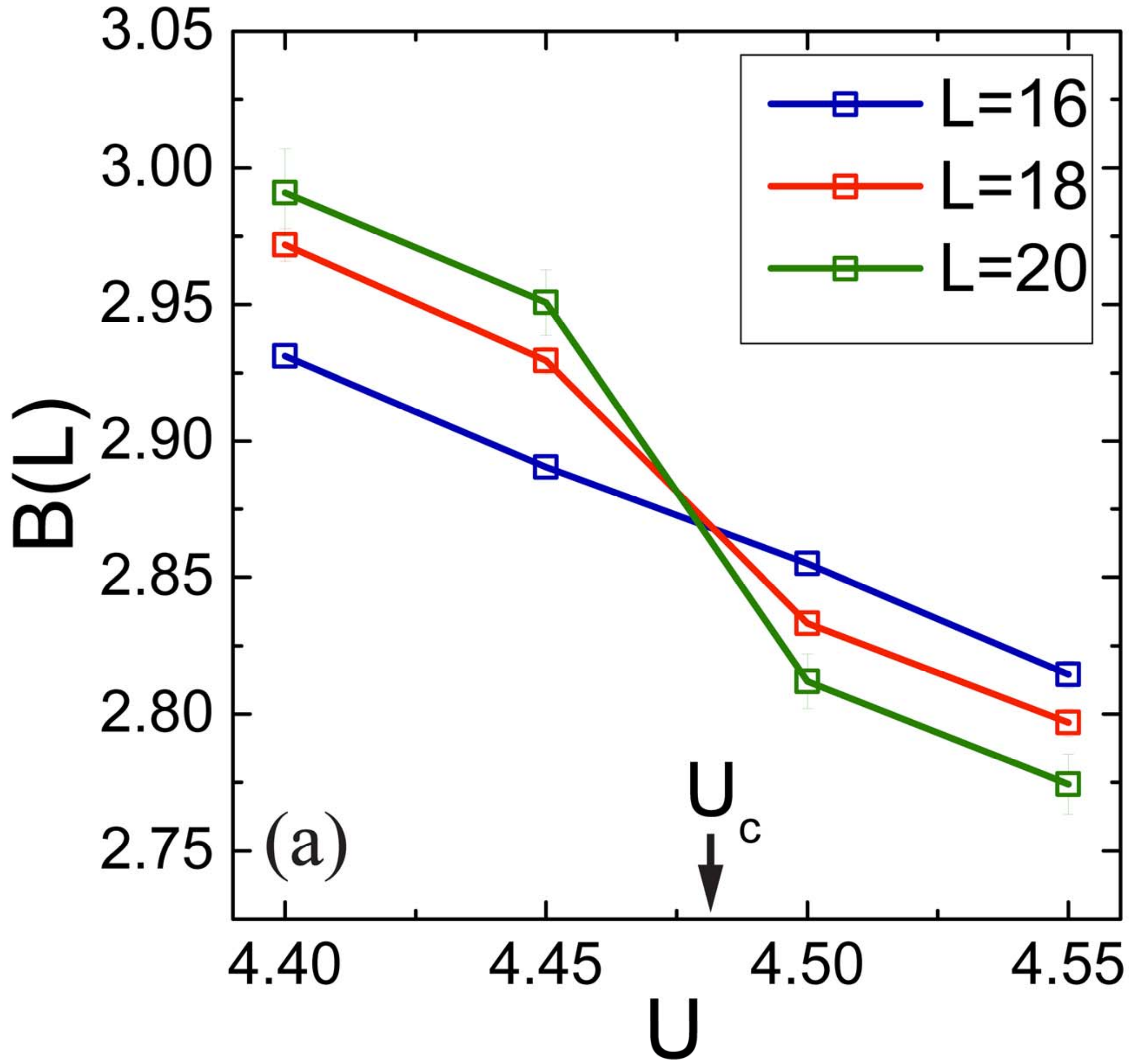}}~~~~~~~~
\subfigure{\includegraphics[height=4.2cm]{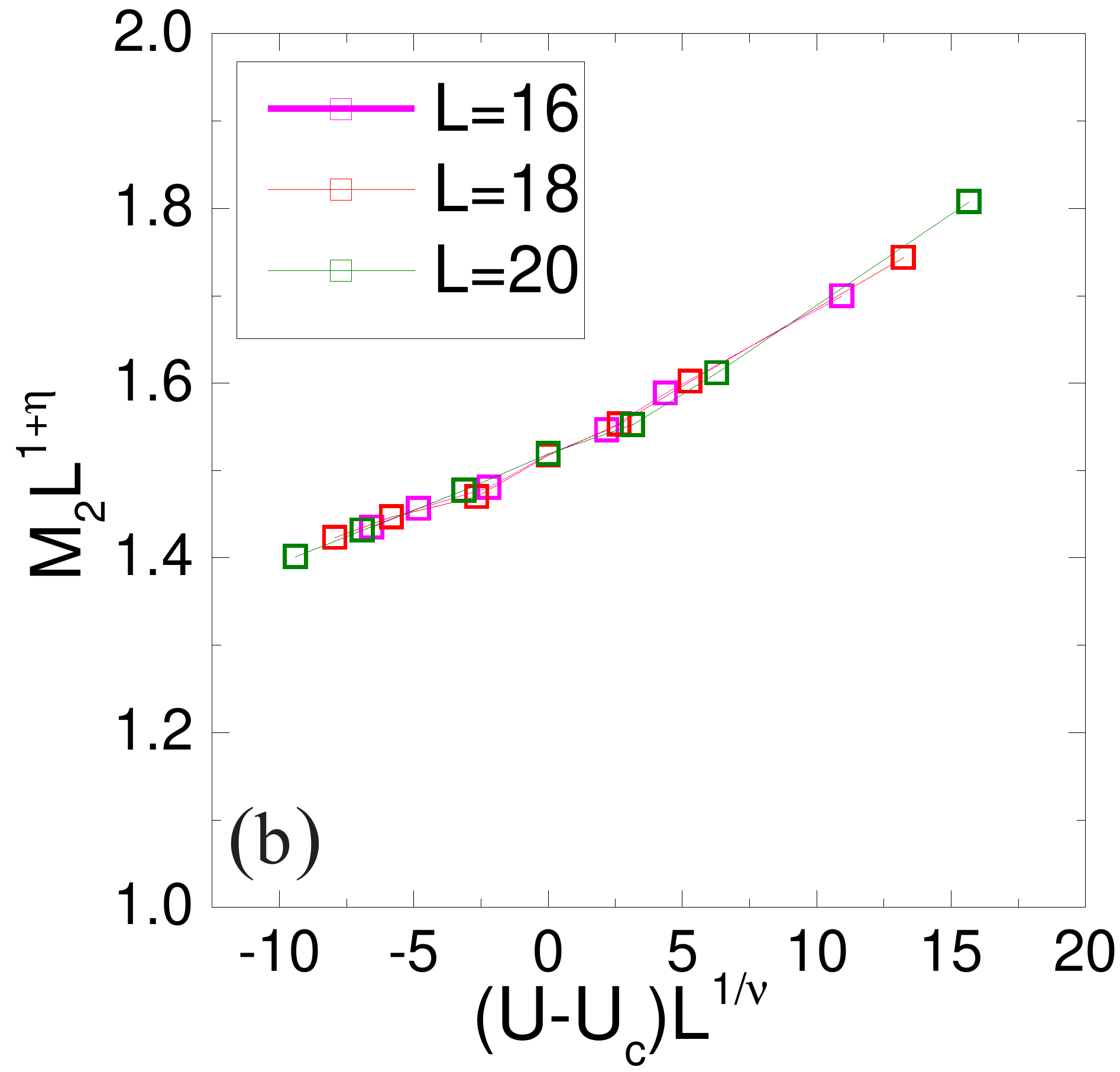}}
\caption{(a) The Majorana QMC results of the interacting topological superconductors of spin-1/2 electrons featuring a topological quantum phase transition. The crossing point of Binder ratio for time-reversal symmetry breaking order parameter shows that the quantum phase transition occurs at $U_c$$\approx$4.48 for $\Delta$=0.3 and $\mu$=$-$0.5. (b) The data collapse analysis of the critical behavior around the time-reversal symmetry breaking transition reveals that  $\nu$$\approx$0.63 and $\eta$$\approx$0.03.} 
\label{SM-spinful}
\end{figure}

We have performed large-scale sign-problem-free Majorana projector QMC simulations of the correlation effect in the interacting topological superconductors described by \Eq{SM-tsc} with $\Delta=0.3t$ and $\mu=-0.5t$ (hereafter we set $t=1$ for simplicity). It is clear that the topological superconductor is stable against weak interaction $U$. When $U$ is strong enough, we expect that the system shall possess a finite singlet pairing, which spontaneously breaks the symmetry $P=T^+T^-=\sigma^z$. From computing the Binder ratio $B(L)$ of the singlet order parameter $\Delta_s=\langle c^\dag_{i\uparrow} c^\dag_{i\downarrow}\rangle$ with size $L\times L$, we can determine the critical values $U_{c}$ of the spontaneous symmetry-breaking, as shown in \Fig{SM-spinful}(a). The quantum critical point $U_c \approx 4.48$ is obtained from the crossing point of Binder ratio of different system sizes. Moreover, in the ordered phase, we found that the singlet pairing amplitude $\Delta_s$ is pure imaginary because the value of $\Delta_s^2$, obtained through the finite-size scaling of the correlation function  $\langle c^\dag_{i\uparrow} c^\dag_{i\downarrow} c^\dag_{j\uparrow} c^\dag_{j\downarrow}\rangle$ with $\vec r_j=\vec r_i+(L/2,L/2)$, is negative. For instance, we obtain $\Delta_s\approx \pm 0.15 i$ for $U=4.6$, indicating that the system spontaneously breaks time-reversal symmetries $T^+$ and $T^-$ in the ordered phase. In other words, the system undergoes a topological quantum phase transition from a topological SC to a topologically-trivial SC which breaks time-reversal symmetry spontaneously. 

Our sign-problem-free QMC can also study the critical behaviors of the topological quantum phase transition with spontaneous time-reversal breaking. From the data collapse analysis, as shown in \Fig{SM-spinful}(b) , we obtain the critical exponents $\nu$$\approx$0.63 and $\eta$$\approx$0.03, which is quite consistent with the Ising quantum critical point in 2+1 dimensions.

\end{widetext}


\begin{thebibliography}{10}

\bibitem{XGWenbook} X.-G. Wen, {\it Quantum Field Theory of Many-body Systems}, (Oxford
    University Press, New York, 2004).

\bibitem{sachdevbook} S. Sachdev, {\it Quantum Phase Transitions}, 2nd ed. (Cambridge University Press, Cambridge, 2011).

\bibitem{Fradkinbook} E. Fradkin, {\it  Field Theories of Condensed Matter Physics}, 2nd ed. (Cambridge University Press, Cambridge, 2013).

\bibitem{kivelson-rmp} S. A. Kivelson, I. P. Bindloss, E. Fradkin, V. Oganesyan, J. M. Tranquada, A. Kapitulnik, and C. Howald, Rev. Mod. Phys. {\bf 75}, 1201 (2003).

\bibitem{xiaogang-rmp} P. A. Lee, N. Nagaosa, and X.-G. Wen, Rev. Mod. Phys. {\bf 78}, 17 (2006).

\bibitem{Blankenbecler-81} R. Blankenbecler, D. J. Scalapino, and R. L. Sugar, Phys. Rev. D {\bf 24}, 2278 (1981).

\bibitem{Hirsch-81} J. E. Hirsch, D. J. Scalapino, R. L. Sugar, and R. Blankenbecler, Phys. Rev. Lett. {\bf 47}, 1628 (1981).

\bibitem{Hirsch-85} J. E. Hirsch, Phys. Rev. B {\bf 31}, 4403 (1985).

\bibitem{HQLin-88} J. E. Hirsch and H. Q. Lin, Phys. Rev. B {\bf 37}, 5070 (1988).

\bibitem{Sugiyama-86} G. Sugiyama and S. Koogin, Annals of Phys. {\bf 168}, 1 (1986).

\bibitem{Sorella-89} S. Sorella, S. Baroni, R. Car and M. Parrinello. Europhys. {\bf 8}, 663 (1989).

\bibitem{White-89} S. R. White, D. J. Scalapino, R. L. Sugar, E. Y. Loh, J.
E. Gubernatis and R. T. Scalettar. Phys. Rev. B, {\bf 40}, 506 (1989).

\bibitem{Ceperley-86} D. Ceperley and B Alder, Science {\bf 231}, 555 (1986).

\bibitem{Shiwei-95} S. Zhang, J. Carlson, and J. E. Gubernatis, Phys. Rev. Lett. {\bf 74}, 3652 (1995).

\bibitem{Prokofev-98} N. V. Prokofev, B. V. Svistunov, and I. S. Tupitsyn, 
    Physics Letters A {\bf 238}, 253 (1998).

\bibitem{Sandvik-99} A. W. Sandvik, Phys. Rev. B {\bf 59}, R14157 (1999).

\bibitem{Foulkes-01} W. M. C. Foulkes, L. Mitas, R. J. Needs, and G. Rajagopal, 
    Rev. Mod. Phys. {\bf 73}, 33 (2001).

\bibitem{Assaad-08} F. F. Assaad and H. G. Evertz, {\it Computational Many Particle Physics, Lecture Notes in Physics} {\bf 739}, 277 (2008).

\bibitem{Gull-11} E. Gull, A. J. Millis, A. I. Lichtenstein, A. N. Rubtsov, M. Troyer, and P. Werner, 
    Rev. Mod. Phys. {\bf 83}, 349 (2011).

\bibitem{Wiese-99} S. Chandrasekharan and U. J. Wiese, Phys. Rev. Lett. {\bf 83}, 3116 (1999).

\bibitem{Suzuki-92} N. Hatano and M. Suzuki, Physics Letters A {\bf 163}, 246 (1992).

\bibitem{Loh-90} E. Y. Loh, J. E. Gubernatis, R. T. Scalettar, S. R. White, D. J. Scalapino, and R. L. Sugar, Phys. Rev. B {\bf 41}, 9301 (1990).

\bibitem{Zaanen-08} J. Zaanen, 
    Science {\bf 319}, 1205 (2008).

\bibitem{Troyer-05} M. Troyer and U.-J. Wiese, Phys. Rev. Lett {\bf 94}, 170201 (2005).

\bibitem{Congjun-05} C. Wu and S.-C. Zhang, Phys. Rev. B, {\bf 71}, 155115 (2005).

\bibitem{LJY-15a} Z.-X. Li, Y.-F. Jiang and H. Yao, Phys. Rev. B {\bf 91}, 241117(R) (2015).

\bibitem{beenakker-13} C. W. J. Beenakker, Annu. Rev. Con. Mat. Phys. {\bf 4}, 113 (2013).

\bibitem{Berg-12} E. Berg, M. A. Metlitski, and S. Sachdev, Science {\bf 338}, 1606 (2012).

\bibitem{Berg-15a} Y. Schattner, S. Lederer, S. A. Kivelson, and E. Berg, Phys. Rev. X {\bf 6}, 031028 (2016).

\bibitem{LWYL-15a} Z.-X. Li, F. Wang, H. Yao, and D.-H. Lee, arXiv:1512.04541 (2015).

\bibitem{Berg-15b} Y. Schattner, M. H. Gerlach, S. Trebst, and E. Berg, Phys. Rev. Lett. {\bf 117}, 097002 (2016).

\bibitem{LWYL-15b} Z.-X. Li, F. Wang, H. Yao, and D.-H. Lee, Science Bulletin {\bf 61}, 925 (2016).

\bibitem{SAK} S. A. Kivelson, Science Bulletin {\bf 61}, 911 (2016).

\bibitem{Ashvin-15} P. T. Dumitrescu, M. Serbyn, R. T. Scalettar, and A. Vishwanath, Phys. Rev. B {\bf 94}, 155127 (2016).

\bibitem{Qi-13} X.-L. Qi, New J. Phys. {\bf 15}, 065002 (2013).

\bibitem{Yao-13} H. Yao and S. Ryu, Phys. Rev. B {\bf 88}, 064507 (2013).

\bibitem{Ryu-12} S. Ryu and S.-C. Zhang, Phys. Rev. B {\bf 85}, 245132 (2012).

\bibitem{JLKY-16} Y.-F. Jiang, Z.-X. Li, S. A. Kivelson, and H. Yao, arXiv:1607.01770.

\bibitem{Shailesh-14} E. F. Huffman and S. Chandrasekharan, Phys. Rev. B {\bf 89}, 111101 (2014).

\bibitem{Lei-15} L. Wang, Y. H. Liu, M. Iazzi, M. Troyer and G. Harcos, Phys. Rev. Lett. {\bf 115}, 250601 (2015).

\bibitem{SM} See the Supplemental Material for the details.

\bibitem{Qi-Zhang-11} X.-L. Qi and S.-C. Zhang, Rev. Mod. Phys. {\bf 83}, 1057 (2011).

\bibitem{Hasan-Kane-10} M. Z. Hasan and C. L. Kane, Rev. Mod. Phys. {\bf 82}, 3045 (2010).

\bibitem{Kitaev-09} A. Kitaev, in {\it Periodic Table for Topological Insulators and Superconductors} (AIP, New York, 2009) [AIP Conf. Proc. {\bf 1134}, 22 (2009)].

\bibitem{Schnyder-08} A. P. Schnyder, S. Ryu, A. Furusaki, and A. W. W. Ludwig, Phys. Rev. B {\bf 78}, 195125 (2008).

\bibitem{Qi-Raghu-09} X.-L. Qi, T. L. Hughes, S. Raghu, and S.-C. Zhang, Phys. Rev. Lett. {\bf 102}, 187001 (2009).

\bibitem{Assaad-11} M. Hohenadler, T. C. Lang, and F. F. Assaad, Phys. Rev. Lett. {\bf 106}, 100403 (2011).

\bibitem{CJWu-11} D. Zheng, G.-M. Zhang, and C. Wu, Phys. Rev. B {\bf 84}, 205121 (2011).

\bibitem{Assaad-12} M. Hohenadler, Z. Y. Meng, T. C. Lang, S. Wessel, A. Muramatsu, and F. F. Assaad, Phys. Rev. B {\bf 85}, 115132 (2012).

\bibitem{Assaad-13} F. F. Assaad, M. Bercx, and M. Hohenadler, Phys. Rev. X {\bf 3}, 011015 (2013).

\bibitem{Fiete-13} H.-H. Hung, L. Wang, Z.-C. Gu, and G. A. Fiete, Phys. Rev. B {\bf 87}, 121113 (2013).

\bibitem{Wessel-13} T. C. Lang, A. M. Essin, V. Gurarie, and S. Wessel, Phys. Rev. B {\bf 87}, 205101 (2013).

\bibitem{CKXu-15}K. Slagle, Y.-Z. You, and C. Xu, Phys. Rev. B {\bf 91}, 115121 (2015).

\bibitem{CKXu-16} Y.-Y. He, H.-Q. Wu, Y.-Z. You, C. Xu, Z. Y. Meng, and Z.-Y. Lu, Phys. Rev. B {\bf 93}, 115150 (2016).

\bibitem{Congjun-16} Z. C. Wei, C. Wu, Y. Li, S. Zhang, and T. Xiang, Phys. Rev. Lett. {\bf 116}, 250601 (2016).

\bibitem{Isakov-06} S. V. Isakov, Y. B. Kim, and A. Paramekanti, Phys. Rev. Lett. 97, 207204 (2006).
\bibitem{Melko-12} S. V. Isakov, M. B. Hastings, and R. G. Melko, Science {\bf 335}, 193 (2012).
\bibitem{FIQCP-15} Z.-X. Li, Y.-F. Jiang, S.-K. Jian, and H. Yao, arXiv:1512.07908 (2015).
\bibitem{WAGuo-16} H. Shao, W. Guo, and A. W. Sandvik, Science {\bf 352}, 213 (2016).


\end{thebibliography}
\end{document}